# Research assessment by percentile-based double rank analysis

**Ricardo Brito**[a*], **Alonso Rodríguez-Navarro**[a,b]

[a] Departamento de Estructura de la Materia, Física Térmica y Electrónica and GISC, Universidad Complutense de Madrid, Plaza de las Ciencias 3, 28040-Madrid, Spain
e-mail address: brito@ucm.es

[b] Departamento de Biotecnología-Biología Vegetal, Universidad Politécnica de Madrid, Avenida Puerta de Hierro 2, 28040-Madrid, Spain
e-mail address: alonso.rodriguez@upm.es

* Corresponding author: Ricardo Brito

**Abstract**

In the double rank analysis of research publications, the local rank position of a country or institution publication is expressed as a function of the world rank position. Excluding some highly or lowly cited publications, the double rank plot fits well with a power law, which can be explained because citations for local and world publications follow lognormal distributions. We report here that the distribution of the number of country or institution publications in world percentiles is a double rank distribution that can be fitted to a power law. Only the data points in high percentiles deviate from it when the local and world $\mu$ parameters of the lognormal distributions are very different. The likelihood of publishing very highly cited papers can be calculated from the power law that can be fitted either to the upper tail of the citation distribution or to the percentile-based double rank distribution. The great advantage of the latter method is that it has universal application, because it is based on all publications and not just on highly cited publications. Furthermore, this method extends the application of the well-established percentile approach to very low percentiles where breakthroughs are reported but paper counts cannot be performed.

# 1. Introduction

Research assessment is an absolute requirement to perform a competent research policy. States and private institutions invest large amounts of funds in research, and society and private investors must know the efficiency of these investments by evaluating research outputs (Martin and Irvine, 1983; Garfield and Welljams-Dorof, 1992; Martin, 1996). In the case of applied research directly focused on the improvement of products or services these outputs have many possibilities of assessment attending to their economic benefits. In contrast, this assessment is much more difficult for basic research. In this case, the assessment can be analyzed in two contexts: the achievement of discoveries and scientific advancements, and the economic benefits as in applied research. However, the latter neither can be easily established nor is the only target of basic research (Salter and Martin, 2001; Bornmann, 2012); even the method that should be applied to this economic analysis is under debate (Abramo and D'Angelo, 2014, 2016; Bornmann and Haunschild, 2016b; Glanzel *et al.*, 2016). Therefore, it seems that the best evaluation of basic research must be done by attending to its scientific achievements. However, even by focusing the assessment of basic research exclusively on these achievements, the assessment is intrinsically difficult because of the intangible nature of the product to be measured (Martin and Irvine, 1983).

Scientific publications are tangible and easily measured. However, although scientific achievements are communicated in publications not all publications communicate real scientific advances. In fact, a large proportion of the published research is "normal science" (Kuhn, 1970) that supports real achievements, but a very low proportion of all publications reports important achievements.

As a consequence of the described needs and difficulties, in the last twenty years, there has been a Cambrian explosion of metrics (van Noorden, 2010) or metric tide (Wilsdon *et al.*, 2015). In this scenario, it has been suggested that no more metrics should be added unless their added value is demonstrated (Waltman, 2016). Many of these metrics are based on the number of publications, but, using a sports simile, counting publications in research is somewhat like counting the kicks in European football rather than counting the goals (Rodriguez-Navarro and Narin, 2017). The weakness of this simile is that football goals are easily recognizable but this easiness does not apply to scientific achievements. Therefore, many metrics and indicators "are based on count what can be easily counted rather than what really counts" (Abramo and D'Angelo, 2014, p. 1130). In fact, 45 years ago, Francis Narin stated that "the relationship between bibliometric measures and other measures may only be validated using a "rule of reason approach" (Narin, 1976, p. 82), which explains the causes for a more recent feeling of Harnad (2009, p. 149): "so we have thus far been rather passive about the validation of our scientific and scholarly performance metrics, taking pot-luck rather than systematically trying to increase their validity, as in psychometrics."

Citation analysis is apparently the solution for grading the importance of results of research, because citation counts seem to correlate with expert assessments (a review of old literature is in Narin, 1976; examples of more recent publications are: Rinia *et al.*, 1998; Aksnes and Taxt, 2004; Allen *et al.*, 2009). However, the debate is still open (MacRoberts and MacRoberts, 1989, 1996; Adler *et al.*, 2009) and the conceptual clarity of citation analysis has been questioned (Martin and Irvine, 1983), because it possibly reflects "impact" or "influence" but the relationship of these concepts with "quality", "importance," or "scientific advance" is less clear. In any case, although citation counts correlate with certain dimensions of research assessment they do not measure it, which implies that it cannot be applied to low aggregation levels: individual researchers or small groups (van-Raan, 2005; Allen *et al.*, 2009; Ruiz-Castillo, 2012).

Another difficulty of citation analysis is the skewed distribution of publications attending to the number of citations (Seglen, 1992; Albarrán *et al.*, 2011a), which makes it difficult to extract relevant information for research assessment from the analysis of simple citation counting. Several approaches have been proposed to extract this information considering citation distribution (Glanzel and Schubert, 1988; Adams *et al.*, 2007; Bornmann *et al.*, 2008; Leydesdorff and Bornmann, 2011; Leydesdorff *et al.*, 2011; Bornmann *et al.*, 2013c; Li *et al.*, 2013; Bornmann and Mutz, 2014; Glanzel *et al.*, 2014; Albarrán *et al.*, 2015; Bornmann and Haunschild, 2016a; Schneider and Costas, 2017), including some that specifically attend to both the number of highly and lowly cited papers (Albarrán *et al.*, 2011c; 2011b; 2011d). All these methods have been developed under strict mathematical and statistical considerations but all have the aforementioned problem of difficult validation.

Citation analysis can be focused on counting the number of highly cited papers, which might give an estimate of the number of important scientific achievements (Martin and Irvine, 1983; Plomp, 1994; Martin, 1996; Tijssen *et al.*, 2002; Aksnes and Sivertsen, 2004; Bonaccorsi, 2007; Rodríguez-Navarro, 2012; Rons, 2013; González-Betancor and Dorta-González, 2017). The simplicity of this idea, however, conceals many difficulties, starting with its own definition: "highly cited," "top-cited," "most frequently cited," etc. (Bornmann, 2014), which implies the arbitrariness of selecting the citation level that should be used (Schreiber, 2013a) and, more importantly, with the question about whether highly cited publications really reflects high scientific influence (Waltman *et al.*, 2013).

Citation counts must be field normalized (Li *et al.*, 2013; Ruiz-Castillo and Waltman, 2015; Waltman, 2016); among the different normalization procedures that can be used there is a method of citation analysis: the percentile rank approach, which intrinsically implies normalization of the citation count data. This approach, which has advantages over other approaches, has been extensively investigated (Bornmann, 2010; Bornmann *et al.*, 2013a; Bornmann *et al.*, 2013b; Waltman and Schreiber, 2013), and allows generating a single measure of citation impact by giving different weights to different percentile rank classes (Bornmann and Mutz, 2011; Leydesdorff and Bornmann, 2011; Leydesdorff *et al.*, 2011; Rousseau, 2012; Bornmann, 2013).

With this same idea of obtaining a single measure of citation impact Rodríguez-Navarro (2011) used a different approach. Firstly, he focused only on the percentiles in the high-citation tail of the citation distribution, assuming that this tail contains the information to estimate the number of important scientific achievements, as described above. Secondly, he did not fix the weights for the percentile rank classes but calculated them through linear regression analysis maximizing the correlation of the single measure with the number of Nobel Prize achievements in several high-level research institutions and advanced countries. The resulting index showed high correlation with the number of Nobel Prize achievements and with the articles published in *Nature* and *Science*. Interestingly, a further study of this approach showed that its success occurred because the upper tails of the citation distributions across countries and institutions do not deviate very much from a power law, independent of whether other functions might explain more accurately tail distribution (Price, 1976; Ruiz-Castillo, 2012; Brzezinski, 2015; Katz, 2016). The power law adjusted to the tail allows estimating the frequency of very highly cited papers or the likelihood of publishing them (Rodríguez-Navarro, 2016).

This finding, however, was more conceptual than useful for research assessment. The difficulty lies in the fact that the proportion of publications that can be treated as a power law in the upper tail can be very low (Ruiz-Castillo, 2012; Brzezinski, 2015). Therefore, an ideal method would be one that produces the same results by using most of the publications and not just those highly cited. This goal was achieved with the double rank analysis, in which the ranking number of a publication in a country or institution is expressed as a function of its world ranking number. The resulting function can be well fitted by a power law, which allows estimating the likelihood of any country or institution producing a very highly cited paper (Rodríguez-Navarro and Brito, 2018; RNB, henceforth).

Therefore, the double rank function for research assessment produces an indicator based on all types of publications, and not just on highly cited publications, and this indicator can be validated in terms of Nobel Prize achievements. However, such an indicator has the intrinsic problem of requiring a specific method of calculation, which is completely different from the well-established procedure of the aforementioned percentile-based research assessment. Fortunately, intuition suggests that a percentile distribution of publication according to the number of citations is in fact a double rank distribution, and therefore, it should be well described by a power law.

Attending to this intuition, this study aimed to find whether the percentile distribution of publications is a double rank distribution that can be well fitted to a power law and used to estimate the likelihood of publishing very highly cited papers. For this purpose this study is divided in four sections. The first section analyzes the percentile double-rank plots in lognormal distributions that have $\mu$ and $\sigma$ values characteristic of citation distributions. The second section compares percentile and normal double rank plots using the data obtained in a previous study (RNB). In the third section, we compare the USA/EU research performance ratios previously obtained by analyzing the high-citation tails (Rodríguez-Navarro, 2016) with those obtained using the percentile-based double rank analysis. Finally, in the fourth section, we demonstrate that the Leiden ranking percentile data (http://www.leidenranking.com/) fit well with power laws.

## 2. Methods

### 2.1. Mathematical modeling

We assumed that the citation distribution obeyed a continuum lognormal function (Redner, 2005; Radicchi *et al.*, 2008; Stringer *et al.*, 2010; Evans *et al.*, 2012; Thelwall and Wilson, 2014a; 2014b), of parameters "$\mu$" and "$\sigma$" that varied within narrow limits. For an institution that publishes $N$ papers, the number of papers that receive between $c$ and $c + dc$ citations is given by the lognormal distribution with the form:

$$f(c; N, \mu, \sigma) = \frac{N}{c\sigma\sqrt{2\pi}} \exp\left[-\frac{(\ln(c) - \mu)^2}{2\sigma^2}\right] dc \qquad [1]$$

which is appropriately normalized, so that the integration from $c = 0$ (no citations) to $c = \infty$ (arbitrarily large number of citations) equals the total number of papers $N$:

$$\int_0^\infty f(c; N, \mu, \sigma) dc = N \qquad [2]$$

In the expressions above, it is implicit that the distribution of citation is a continuum variable, instead of a discrete variable, as it is the case in real citation counts. However, we assume a continuum citation variable as the mathematical analysis become simpler than in the case of a discrete variable (Li *et al.*, 2013), but it is also possible to use discrete variables.

For the percentile analysis, two distributions are required: A first one for all the papers published in the world in the studied area, and a second one for the papers published by the institution. Therefore, we need two sets of parameters (N, $\mu$, $\sigma$), one for the world and a second one for the institution or country. The parameters for the world will be denoted by the subindex "w". The parameters are determined by using a two- or three-year counting window for the world and for countries and institutions that represent the highest level of scientific performance and a reasonable minimum. For comparisons, we use the same parameters as in our previous study (RNB; Table 2); they are recorded in figures.

*2.2. Retrieval and counting methods in section 3.2*

We applied the percentile-based double rank analysis to the data obtained in a previous study (RNB), in which the methods for paper retrieval and counting are described. Briefly, data were obtained from the Web of Science using the "Advanced search" feature in the research areas (SU=) of Plant Sciences and Physiology, and topic (TS=) of graphene, and for the research countries (CU=) and years (PY=) recorded in each case. In all cases we counted domestic publications. The retrieved publications were ordered by using the WoS feature "Times cited - - highest to lowest" and downloaded by using the WoS feature "Create Citation Report."

To determine the number of publications in each percentile, we ordered the world publications in a field and year from the highest to lowest number of citations and counted the number of publications in each percentile ($x\%$), starting at the top of the list (the $x\%$ percentile contains the top $x \cdot N/100$ papers rounded to the closest integer; $N$ includes the publications with cero citations). In this type of apportionment, the paper in the $x \cdot N/100$ position may be in the middle of several publications with the same number of citations both in the world and country or institution lists. This is a significant problem for the calculation of some percentile indicators (Schreiber, 2013b). Here, after ordering the publications by their number of citations in the recorded period (e.g. two-year citation window), tied publications remained ordered by the total number of citations recorded in the database at the moment of the search; the world $x\%$ set was constituted by the top $x \cdot N/100$ papers. In this set of publications we counted the number of those corresponding to the investigated country.

*2.3. Retrieval and counting methods in section 3.3*

The percentile distribution of publications in the research areas (SU=) of Chemistry, Physics, and Biochemistry & Molecular biology OR Microbiology were obtained using the features provided by the WoS and the total number of citations from the publication year either 2006 or 2007 up to the date of the search, August 15, 2017. We retrieved the papers for the world, the USA, and the EU using the "advanced search" feature as previously described (Rodríguez-Navarro, 2016), and the retrieved papers were ordered from the highest to the lowest number of citations. To calculate the percentile data without downloading hundreds of thousands of papers, we determined the number of papers corresponding to each percentile in the world set ($N_x = x \cdot N/100$), for each percentile we recorded the number of citations of the paper with the rank number corresponding to the number of papers ($N_x$) in each percentile. Because in some percentiles the number of citations was repeated in many papers, we also recorded the rank numbers of the first and last papers with the same number of citations. Then, in the USA and EU paper sets, we determined the number of papers included in each percentile of the world set by first using the number of citations found in the world set.

Because in some cases many papers had this number of citations we again recorded the rank numbers of the first and last papers with this number of citations. Then, we fixed the number of papers in each *x*% percentile by adding to the ranking number of the first paper with the corresponding number of citations a number of papers that was equal to the USA or EU/world ratio of the numbers of tied papers. This method assumes that considering a certain number of citations, the tied USA and EU papers are homogeneously distributed among the tied world papers (e.g. USA papers are not ordered first in the world list). We used this method because it is a simple formal solution to the problem of papers with the same citation counts (Schreiber, 2013b). However, we found that the inclusion of all papers with the same number of citations in the same percentile would have not affected the results of this study. We recorded 12 percentiles: 100, 50, 30, 20, 10, 8, 5, 3, 2, 1, 0.5, and 0.2.

We compared our results to previous ones obtained from the analysis of the upper tail (Rodríguez-Navarro, 2016). For two cases, "Chemistry" and "Physics," the searches were identical to the previous ones but in one case, the research area of (SU=) "Biochemistry & Molecular Biology" was complemented with the research area of "Microbiology." We added "Microbiology" to have a more similar number of papers in the three areas of study. This addition does not affect the comparison of present and previous results, because the USA/EU performance ratios in "Biochemistry & Molecular Biology" and "Microbiology" are similar (the same conclusion might not be reached from the data in Herranz and Ruiz-Castillo, 2011, but the treatment of the data in Herranz and Ruiz-Castillo, 2011 and Rodríguez-Navarro, 2016 are different). To compare these results it was necessary to know the percentile that corresponds with a Nobel Prize-level publication. For this purpose, we assumed that the USA obtained 1.1, 0.9, and 0.9 Nobel Prize achievements per year in chemistry, physics, and biology (Rodríguez-Navarro, 2016) and operated as explained below when these data were applied.

*2.4. Fits of Leiden ranking data*

The purpose of section *3.4* of our study was to calculate the goodness of fit of the Leiden indicators: P, $P_{top50\%}$, $P_{top10\%}$, and $P_{top1\%}$ to a power law function. For this purpose we used the Leiden Ranking 2017 (http://www.leidenranking.com/ranking/2017/list), fractional counting in the fields of "biomedical and health sciences," "life and earth sciences," and "physical sciences and engineering," and in the time periods: 2012-2015, 2011-2014, 2010-2013, and 2009-2012, which gave 12 fittings for each university. Then we selected six universities in the Leiden ranking with the condition that the $P_{top1\%}$ indicator, calculated using fractional counting, was as variable as possible, but having a minimum value of approximately 10 to avoid that the variability of this indicator had a strong influence on our fitting calculations. Because in all these universities a high research level could be expected we added two universities in which P$_{top1\%}$ could be as low as 4. Attending to these conditions we selected the following universities: MIT, Toronto, Duke, Queensland, Copenhagen, Ghent, Milano, Barcelona, Porto, and Bath.

*2.5. Fits of power laws*

The fits of power laws to empirical data present some difficulties and, in some cases, least-squares fittings to log transformed data are not recommended (Clauset *et al*., 2009). However, our case was special for three reasons: (i) we did not observe fluctuations in the tails (Fig. 1), (ii) different fitting methods give different weights to high and low percentile data and we had to select the method that give more weight to low percentiles, where we had maximum interest, (iii) both the log-log plots (e.g., Fig 1) and $R^2$ values (20-0.2% in Table 1) showed that we were dealing with real power laws. Attending to these considerations, we used least-square fittings to log-transformed data. The $R^2$ values were obtained comparing the empirical data to those obtained from the fitted equation.

## 3. Results

*3.1. Mathematical percentile-based double rank plots from lognormal distributions*

We hypothesized that the plot of the number of publications in percentiles is a double rank plot as defined previously (RNB). The rationale of this hypothesis is that in the double rank plot the rank position in the y-axis is also the number of publications that exceed a certain number of citations. Now if we plot the number of publications of a country that are contained in the world x% percentile, we are really plotting the number of country publications that exceed the number of citations fixed by the publication in position *x·N*/100 in the world, which means a double rank plot.

In order to construct the percentile double rank plot, first we need to determine the number of citations required to be in the top *x*%. That is, in the top *x*% enter all the publication that get *more* than a threshold amount of citations, which we denote by $c_0$. Assuming that the citation counts follow a lognormal distribution, this number $c_0$ is calculated by solving the equation for the world parameters ($N_w, \mu_w, \sigma_w$):

$$\int_{c_0}^{\infty} f(c; N_w, \mu_w, \sigma_w) dc = \frac{xN_w}{100} \qquad [3]$$

The quantity $c_0$, as defined above, is the number of citation that a paper requires to be in the top *x*%. The integration of equation [3] gives an analytical expression for $c_0$:

$$c_0 = \exp[\mu_w - \sqrt{2}\, \sigma_w \text{Erf}^{-1}(2x/100 - 1)] \qquad [4]$$

where Erf is the error function (Gautschi, 1965). Once $c_0$ is determined, we can calculate the number of papers of an institution in the top *x*% as:

$$N(x) = \int_{c_0}^{\infty} f(c; N, \mu, \sigma) dc = \frac{N}{2}\left[1 + \text{Erf}\left(\frac{\mu - \ln(c_0)}{\sqrt{2}\sigma}\right)\right] \qquad [5]$$

Applying this procedure we calculated the number of papers as a function of the percentile and constructed the percentile-based double rank plots. As described above (section 2.1), the simulation of a double rank plot requires two series of simulated citations: one for the world and another for the country or institution. We used four series that simulated country or institution publication (series parameters are given in the figure), which were assembled with a series of data points that simulated the world publications ($N_w$ = 150,000; $\mu_w$ = 1.7; $\sigma_w$ = 1.0). Fig. 1 shows that the resulting log-log plots deviated very little from straight lines. Maximum deviation (Fig. 1a) occurred in the series with the highest $\mu$ parameter, which simulated the most efficient research institution (e.g., Massachusetts Institute of Technology). Fig. 2 contrasts the percentile-based double rank plots for two simulated institutions, in which the $\mu$ parameter of the institution with the higher number of publications is lower than in the other. This simulation shows that if research is evaluated by measuring the number of papers at a high percentile, the institution with the largest production performs better. However, the advantage of the mass production of papers is lost in benefit of high quality papers if the evaluation is taken at high citation levels (at a percentile x < 0.5% in this case). The red squares in Fig. 2 correspond to an institution with the same parameters $\mu$ and $\sigma$ than those of the simulated world publications. In such case, Eq. [5] can be solved analytically giving a power law of exponent 1, which is a straight line.

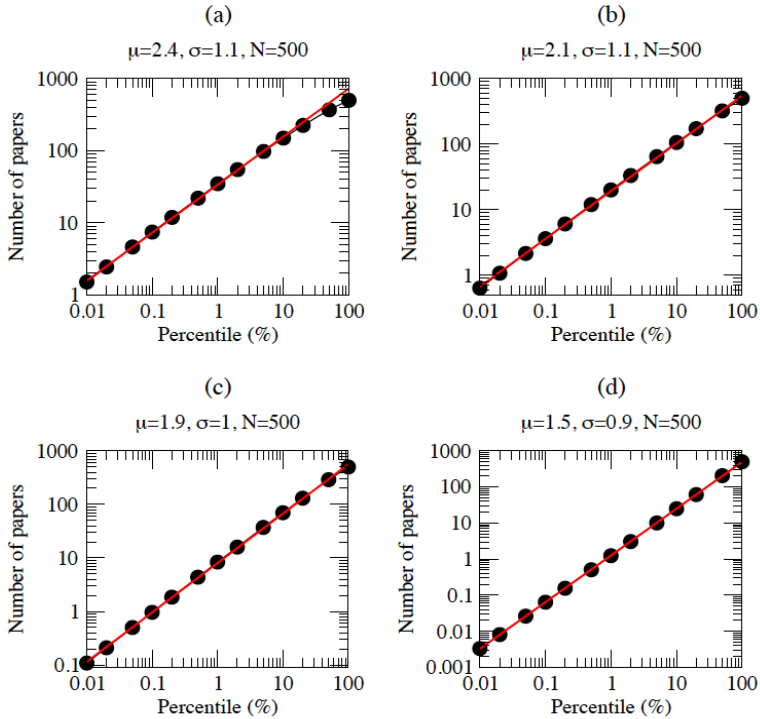

Fig. 1. Percentile-based double rank distributions of the data points from four lognormal distributions. The plots show the percentile positions of the data points of four lognormal distributions, which simulate the publications of four research institutions, with reference to a more numerous lognormal distribution, which simulates the world publications. The parameters of the lognormal distribution that simulated world publications are $N$ = 150,000, $\mu$=1.7 and $\sigma$=1. Straight lines are fittings to power laws to the complete set of points, except the panel (a) where the power law is fitted for percentiles x smaller than 10%. Fitting of the plot data points (without log transformation) to power laws produced the following equations: (a) $N = 33.89\, x^{0.666}$; (b) $N = 19.20\, x^{0.728}$; (c) $N = 8.03\, x^{0.923}$; (d) $N = 1.27\, x^{1.295}$

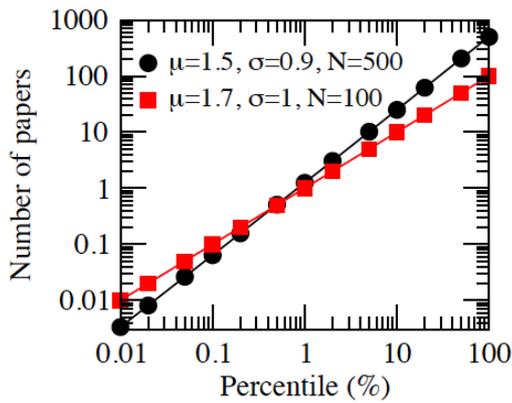

*Fig. 2. Comparison of the percentile-based double rank distributions of the data points from two lognormal distributions. The plots show the percentile positions of the data points of two lognormal distributions, which simulate the publications of two research institutions, with reference to a more numerous lognormal distribution, which simulates the world publications. Symbols: circles, N= 500, μ=1.5, σ=0.9; squares, N = 100, μ = 1.7, σ = 1. The parameters for the world distribution are N = 150,000, μ = 1.7, σ = 1. Fitting of the data points to a power law produced the following equations: circles, $N = 1.27\ x^{1.295}$; squares, $N = 1.0\ x^{1.0}$*

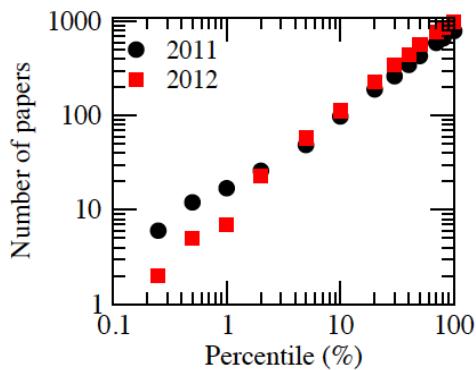

*Fig. 3. Percentile distribution of USA publications on graphene in 2011 and 2012.*

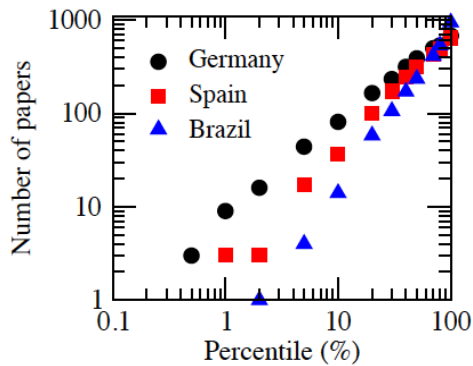

*Fig. 4. Percentile distribution of domestic articles from Germany, Spain, and Brazil published in 2012 in the WoS research area of "Plant sciences." Fitting of the data to a power law produced the following equations: $N_{ger} = 7.84\ x^{0.99}$, $N_{sp} = 2.12\ x^{1.26}$, and $N_{bra} = 2.73\ x^{1.74}$.*

*3.2. Percentile-based versus normal double rank plots*

Next, we constructed the percentile-based double rank plots with the same empirical data that we used previously to describe the normal double rank plots (RNB): publications of several countries in graphene and plant sciences. In all cases, the resulting plots deviated very little from straight lines and the data fitted well with power laws with $R^2$ values higher than 0.98; the percentile-based double rank plots showed a high similarity with our previous double rank plots. For example, comparing Fig. 3 to Fig. 5 in (RNB), both constructed with the same data, similarities are evident; the obvious deviations in the lower part of the plots are softened in Fig. 3. The slight differences in the upper part of the plots might be explained, because, in the present study but not in the previous one, the publications with zero citations were included.

More interesting is the comparison of Fig. 4 of the present study to Fig. 3 in (RNB). The studied data correspond to the publications in 2012 in the WoS research area of "plant sciences" from Germany, Spain, and Brazil. As described above for Fig. 3, deviations from a straight line in the log-log plots were softened in the percentile-based plot. Because of deviations and variability in the normal double rank plots, the lower 30 points were omitted. In contrast, in the percentile-based double rank analysis, all points fitted well with a power law (fitting results are shown in the legend for Fig. 4). In (RNB) the likelihood of Germany, Spain, and Brazil to publish the most cited paper was 0.06, 0.007, and $9.0 \cdot 10^{-5}$, respectively. To compare these likelihoods with those obtained from the percentile distribution, it was necessary to use the equivalence between the most cited paper—first in the rank—and a certain percentile, which implies a simple calculation, considering the total number of papers published. This number was 17,501, and the equivalent percentile was 0.0057%. Using this value of $x$ in the equations given in Fig. 4, the likelihood was 0.05, 0.003, and $3.3 \cdot 10^{-5}$, to be compared with the figures above. Considering that, in our previous calculations, we omitted the lower 30 points and the publications with zero citations, the similarities are evident.

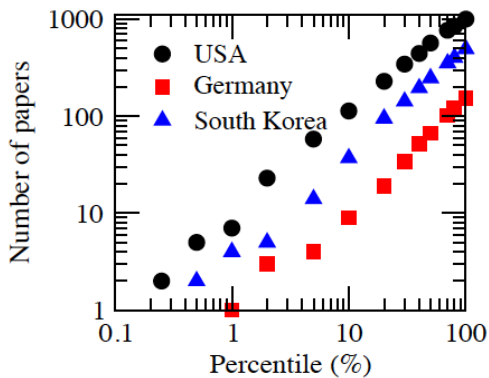

Fig. 5. Percentile distribution of domestic articles published in 2012 from the USA, Germany, and South Korea in the research topic of graphene. Fitting of the data to a power law produced the following equations: $N_{usa} = 9.39\ x^{1.04}$, $N_{ger} = 0.952\ x^{1.08}$, and $N_{sk} = 3.32\ x^{1.09}$.

The comparison of the plots in Fig. 5 to those for graphene in Fig .4 in (RNB) reveals that the double rank plots and percentile distributions are also very similar. In this case, the likelihood to publish the most cited paper was not reported; we calculated them now and they are 0.189, 0.030, and 0.0019 for the USA, South Korea, and Germany, respectively. Similar calculations for the percentile distributions were 0.098, 0.029, and 0.0083, respectively. Again the similarities are evident although there are some differences, that can be explained by the fact that extrapolations are very sensitive to fluctuations.

*3.3. Comparison of percentile-based double rank and upper tail assessments*

To further test the percentile-based double rank analysis, we compared this method of assessment to the previously used assessment of the differences between the USA and EU based on the high-citation tail (Rodríguez-Navarro, 2016). For this purpose we generated the percentile-based plots for the publications in the WoS research fields of Chemistry, Physics, and Biochemistry & Molecular Biology and Microbiology in 2007 and 2006. Fig. 6-8 show that the plots for 2007 in log-log scales are almost perfect straight lines.

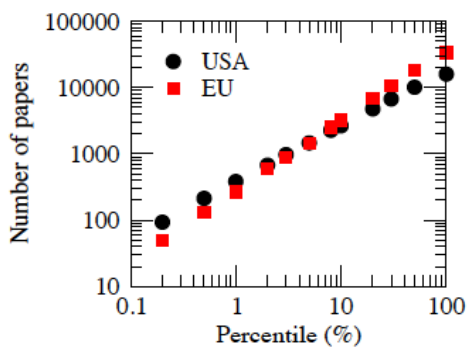

Fig. 6. Percentile distribution of domestic articles published in 2007 from the USA and the EU in the research area of chemistry.

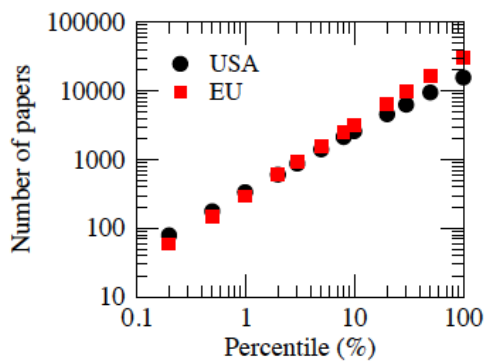

Fig. 7. Percentile distribution of domestic articles published in 2007 from the USA and the EU in the research area of physics.

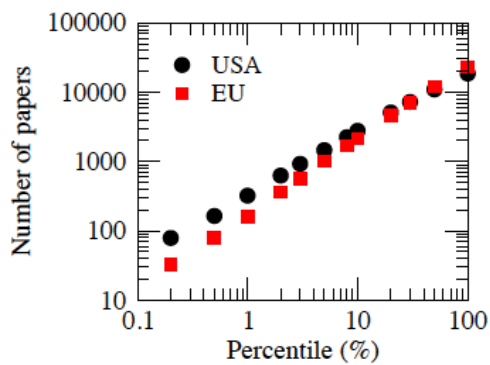

Fig. 8. Percentile distribution of domestic articles published in 2007 from the USA and the EU in the research areas of biochemistry & molecular biology and microbiology.

However, we have shown in section 3.1 that, when $\mu$ and $\sigma$ of the lognormal distribution of papers by citations of a country or institution differ significantly from the $\mu$ and $\sigma$ of the world lognormal distribution, the percentile-based double rank plot showed a slight deviation from the power law in the upper part of the plot (Fig. 1a).

These parameter differences occur in our comparison of the USA and EU. The $\mu$ values in the EU and world are similar but in the USA they are higher. Consistent with this notion, fitting all the USA data points with a single power law, from 100% to 0.2% (not shown), produced a quite high $R^2$ value (> 0.9) but fitting the data in the 20-0.2% interval produced a higher $R^2$ value (> 0.998). In the EU these differences did not occurred. This suggests that the power law is a good model in the whole percentile interval for the EU but that in the USA the model is better in the 20-0.2% interval. To reveal more clearly these differences between the fittings for the USA and EU, Table 1 records the fittings in two percentile intervals, 20-0.2% and 100-5% showing that they produce different power laws in the USA but not so different in the EU. As a consequence of the differences in the $\mu$ values, the exponents of the double rank power laws were around 0.87 in the USA and 1.0 in the EU (i.e., the plot of the untransformed data is close to a straight line, which indicates that the EU and world research performances are similar).

Table 1. Fittings of the percentile-based double rank plots to power laws for the publications from USA and EU in the fields of chemistry, physics, and biology. The log-log transformed data are presented in Fig. 4-6. Fittings were performed with the data in the 20-0.2% and 100-5% percentile ranges

| Actor-Field[a] | Year | Percentile interval | | | | | |
|---|---|---|---|---|---|---|---|
| | | 20-0.2% | | | 100-5% | | |
| | | A | $\alpha$ | $R^2$ | A | $\alpha$ | $R^2$ |
| USA-Chemistry | 2007 | 376.42 | 0.8522 | 0.9995 | 418.88 | 0.8044 | 0.9919 |
| | 2006 | 364.11 | 0.8769 | 0.9999 | 421.60 | 0.8131 | 0.9881 |
| USA-Physics | 2007 | 330.94 | 0.8859 | 0.9997 | 389.44 | 0.8052 | 0.9970 |
| | 2006 | 342.15 | 0.8806 | 0.9992 | 394.03 | 0.8179 | 0.9952 |
| USA-Biology | 2007 | 330.15 | 0.9200 | 0.9997 | 385.40 | 0.8526 | 0.9958 |
| | 2006 | 304.99 | 0.9614 | 0.9989 | 372.67 | 0.8613 | 0.9949 |
| EU-Chemistry | 2007 | 272.83 | 1.0689 | 0.9996 | 280.94 | 1.0552 | 0.9890 |
| | 2006 | 287.02 | 1.0394 | 0.9989 | 2.93.36 | 1.0373 | 0.9937 |
| EU-Physics | 2007 | 301.23 | 1.0217 | 0.9997 | 316.82 | 1.0017 | 0.9966 |
| | 2006 | 291.27 | 1.0344 | 0.9992 | 318.39 | 1.0038 | 0.9967 |
| EU-Biology | 2007 | 174.51 | 1.0850 | 0.9997 | 191.41 | 1.0518 | 0.9966 |
| | 2006 | 176.96 | 1.0677 | 0.9986 | 186.75 | 1.0541 | 0.9975 |

[a] WoS research areas (SU=) of "Chemistry," "Physics," and "Biochemistry & Molecular Biology" plus "Microbiology"

Next, we compared the percentile-based double rank approach to a previous approach that uses the publications in the power law tail to calculate the frequency of Nobel Prize-level publications (Rodríguez-Navarro, 2016; Rodriguez-Navarro and Narin, 2017). For this purpose, we had to determine the Nobel Prize-level percentile and this was calculated by using the number of publications from the USA and assuming that that the USA obtained 1.1, 0.9, and 0.9 Nobel Prize achievement per year in chemistry, physics, and biology (Rodríguez-Navarro, 2016; percentiles are recorded in Table 2). With the percentile data and the power law equations defined in Table 1 we calculated the theoretic frequencies of Nobel Price level papers for the USA and the EU. Using these data, Table 2 summarizes the USA/EU performance ratios calculated fitting the power law equations into the 20-0.2% range, in which fitting was more accurate.

Table 2. Estimation of the USA/EU research performance ratios, expressed as the ratio of the theoretic frequencies of publishing a Nobel-prize level paper. Comparison of percentile-based double rank calculations in the 20-0.2% percentile interval (see Table 1) with previous calculations fitting the upper tail data points to a power law.

| Field | Year | Percentile-based method | | Power-law tail method[b] | Comparison of methods[c] |
|---|---|---|---|---|---|
| | | Nobel-level %[a] | Calculated ratio | | |
| Chemistry | 2007 | 0.00101 | 6.16 | 3.16 | 1.95 |
| | 2006 | 0.00127 | 3.75 | 3.41 | 1.01 |
| Physics | 2007 | 0.00127 | 2.72 | 1.94 | 1.40 |
| | 2006 | 0.00118 | 3.32 | 2.30 | 1.44 |
| Biology | 2007 | 0.00163 | 5.46 | 3.27 | 1.67 |
| | 2006 | 0.00193 | 3.28 | 2.92 | 1.12 |

[a] Nobel-prize level percentile calculated assuming that USA obtained 1.1, 0.9, and 0.9 Nobel Prize achievement per year in Chemistry, Physics, and Biology (Rodríguez-Navarro, 2016). These data were used to calculate the percentile by using the parameters given in Table 1.
[b] Data taken from Rodríguez-Navarro, 2016; Rodríguez-Navarro and Narin, 2017
[c] Percentile-based research performance ratio divided by upper tail research performance ratio

The results obtained by the percentile-based method were 1.4 times higher than those previously reported using the upper tail method (Rodríguez-Navarro, 2016; Rodriguez-Navarro and Narin, 2017). Then, we determined that at the 0.01% percentile, the percentile-based double rank and the power-law tail produced the same results.

*3.4. Fitting of the Leiden data to power laws*

The results presented so far corresponded to domestic papers and a high level of aggregation, the EU, and single countries. Therefore, we wondered whether, at a lower level of aggregation (e.g., institutions), fractional counting, and more complex elaboration of the research areas, the percentile-based double rank plot could be still well fitted to a power law. For this purpose the data for the Leiden ranking (Waltman *et al.*, 2012; Mutz and Daniel, 2015) offered an excellent opportunity. The ranking only provides data for four percentiles: P, $P_{top50\%}$, $P_{top10\%}$, and $P_{top1\%}$, but for many universities, seven time periods, and three research fields in natural sciences. Even eliminating the universities in which the $P_{top1\%}$ is too low, the number of universities that can be tested is very high. We performed 132 tests with eleven universities, four time periods, and in the three natural sciences fields (Table 3). In nine universities, the $P_{top1\%}$ was always higher than 10 or slightly lower. In the other two universities, Bath and National Autonomous of Mexico, in some cases the $P_{top1\%}$ was 4; the total number of publications was much higher in the latter than in the former.

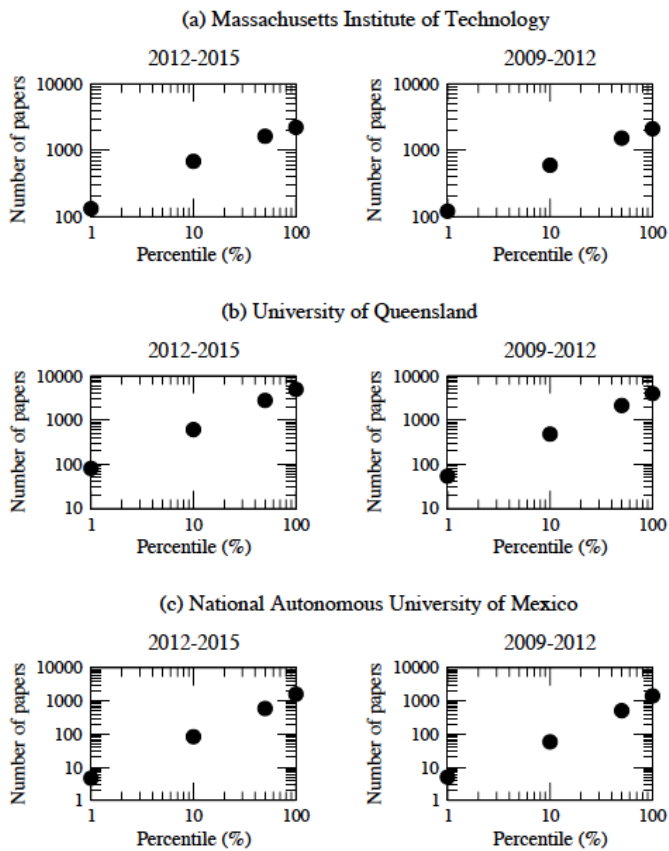

Fig. 9. Double log plot of the Leiden ranking data for three universities in two evaluation periods in the field of "biomedical and health sciences."

Visually, the log-log plots of the percentile-based double rank data were straight lines. Fig. 9 shows the plots of three universities in the Leiden field of "biomedical and health sciences," with power law exponents ranging from 0.6 in the Massachusetts Institute of Technology to 1.4 in the National Autonomous University of Mexico. Consequently, fitting to power laws showed high $R^2$ values. These values were slightly lower in the universities with the lowest values of $\alpha$, Massachusetts Institute of Technology and Duke University, than in the rest of the universities: 0.97 versus 0.99, approximately (Table 3).

Table 3. Fitting to power laws of the P, $P_{top\ 50\%}$, $P_{top\ 10\%}$ and $P_{top\ 1\%}$ indicators in a sample of universities from the Leiden ranking 2017

| | Period | Biomedical and health sciences | | | Life and earth sciences | | | Physical sciences and engineering | | |
|---|---|---|---|---|---|---|---|---|---|---|
| | | A | α | $R^2$ | A | α | $R^2$ | A | α | $R^2$ |
| MIT | 2012-2015 | 141.5 | 0.616 | 0.9776 | 53.2 | 0.670 | 0.9845 | 197.3 | 0.720 | 0.9745 |
| MIT | 2011-2014 | 150.0 | 0.604 | 0.9840 | 53.5 | 0.660 | 0.9841 | 184.0 | 0.737 | 0.9663 |
| MIT | 2010-2013 | 137.0 | 0.618 | 0.9844 | 47.9 | 0.669 | 0.9823 | 179.5 | 0.738 | 0.9628 |
| MIT | 2009-2012 | 137.0 | 0.618 | 0.9844 | 36.4 | 0.724 | 0.9638 | 170.8 | 0.739 | 0.9614 |
| Duke U. | 2012-2015 | 120.7 | 0.918 | 0.9797 | 29.1 | 0.816 | 0.9746 | 23.5 | 0.855 | 0.9831 |
| Duke U. | 2011-2014 | 125.1 | 0.906 | 0.9838 | 30.1 | 0.803 | 0.9713 | 24.1 | 0.856 | 0.9852 |
| Duke U. | 2010-2013 | 124.1 | 0.899 | 0.9869 | 34.7 | 0.765 | 0.9772 | 23.1 | 0.867 | 0.9800 |
| Duke U. | 2009-2012 | 128.1 | 0.883 | 0.9886 | 34.6 | 0.756 | 0.9752 | 22.6 | 0.879 | 0.9685 |
| U. Queensland | 2012-2015 | 79.4 | 0.899 | 0.9990 | 34.4 | 0.928 | 0.9835 | 32.1 | 0.926 | 0.9873 |
| U. Queensland | 2011-2014 | 70.0 | 0.918 | 0.9984 | 31.1 | 0.933 | 0.9806 | 31.9 | 0.915 | 0.9877 |
| U. Queensland | 2010-2013 | 60.6 | 0.932 | 0.9983 | 28.9 | 0.939 | 0.9824 | 32.2 | 0.894 | 0.9855 |
| U. Queensland | 2009-2012 | 54.7 | 0.934 | 0.9994 | 28.4 | 0.920 | 0.9869 | 32.7 | 0.869 | 0.9952 |
| U. Toronto | 2012-2015 | 202.9 | 0.932 | 0.9917 | 22.4 | 0.939 | 0.9834 | 60.4 | 0.859 | 0.9911 |
| U. Toronto | 2011-2014 | 192.2 | 0.940 | 0.9913 | 23.9 | 0.911 | 0.9919 | 64.4 | 0.838 | 0.9948 |
| U. Toronto | 2010-2013 | 186.2 | 0.942 | 0.9900 | 25.5 | 0.895 | 0.9906 | 58.1 | 0.860 | 0.9911 |
| U. Toronto | 2009-2012 | 179.5 | 0.940 | 0.9908 | 22.2 | 0.921 | 0.9828 | 57.4 | 0.855 | 0.9906 |
| U. Copenhagen | 2012-2015 | 89.0 | 0.944 | 0.9948 | 33.1 | 0.924 | 0.9940 | 15.1 | 0.975 | 0.9843 |
| U. Copenhagen | 2011-2014 | 86.3 | 0.934 | 0.9935 | 30.6 | 0.935 | 0.9920 | 15.3 | 0.959 | 0.9932 |
| U. Copenhagen | 2010-2013 | 75.6 | 0.942 | 0.9943 | 29.3 | 0.931 | 0.9933 | 15.9 | 0.940 | 0.9953 |
| U. Copenhagen | 2009-2012 | 63.9 | 0.960 | 0.9928 | 28.6 | 0.925 | 0.9962 | 18.4 | 0.897 | 0.9970 |
| Ghent U. | 2012-2015 | 50.5 | 0.933 | 0.9948 | 36.5 | 0.895 | 0.9971 | 31.6 | 0.908 | 0.9891 |
| Ghent U. | 2011-2014 | 50.9 | 0.933 | 0.9933 | 35.6 | 0.895 | 0.9984 | 31.8 | 0.893 | 0.9957 |
| Ghent U. | 2010-2013 | 41.7 | 0.971 | 0.9903 | 32.3 | 0.911 | 0.9953 | 29.8 | 0.892 | 0.9967 |
| Ghent U. | 2009-2012 | 38.8 | 0.972 | 0.9934 | 27.7 | 0.945 | 0.9895 | 29.2 | 0.884 | 0.9975 |
| U. Milan | 2012-2015 | 33.1 | 1.052 | 0.9973 | 7.6 | 1.066 | 0.9985 | 4.3 | 1.185 | 0.9870 |
| U. Milan | 2011-2014 | 33.8 | 1.039 | 0.9982 | 8.2 | 1.034 | 0.9995 | 4.2 | 1.166 | 0.9932 |
| U. Milan | 2010-2013 | 34.9 | 1.022 | 0.9997 | 7.0 | 1.071 | 0.9969 | 6.2 | 1.059 | 0.9999 |
| U. Milan | 2009-2012 | 30.7 | 1.047 | 0.9984 | 4.3 | 1.185 | 0.9870 | 6.8 | 1.043 | 0.9978 |
| U. Barcelona | 2012-2015 | 37.7 | 0.935 | 0.9982 | 11.0 | 0.999 | 0.9950 | 15.7 | 0.991 | 0.9991 |
| U. Barcelona | 2011-2014 | 31.8 | 0.972 | 0.9945 | 9.2 | 1.048 | 0.9882 | 14.6 | 1.019 | 0.9957 |
| U. Barcelona | 2010-2013 | 27.2 | 0.990 | 0.9907 | 10.1 | 1.020 | 0.9892 | 14.3 | 1.023 | 0.9947 |
| U. Barcelona | 2009-2012 | 22.8 | 1.031 | 0.9878 | 10.5 | 1.005 | 0.9912 | 13.7 | 1.041 | 0.9890 |
| U. Porto | 2012-2015 | 15.1 | 1.095 | 0.9999 | 11.2 | 1.013 | 0.9980 | 16.6 | 1.006 | 0.9931 |
| U. Porto | 2011-2014 | 15.6 | 1.061 | 0.9988 | 8.5 | 1.067 | 0.9938 | 17.3 | 0.993 | 0.9952 |
| U. Porto | 2010-2013 | 11.7 | 1.100 | 0.9997 | 9.1 | 1.013 | 0.9983 | 14.6 | 1.035 | 0.9840 |
| U. Porto | 2009-2012 | 12.1 | 1.052 | 0.9929 | 10.0 | 0.982 | 0.9985 | 15.7 | 0.990 | 0.9958 |
| U. Bath | 2012-2015 | 4.3 | 1.052 | 0.9937 | 4.3 | 0.806 | 0.9991 | 17.1 | 0.867 | 0.9994 |
| U. Bath | 2011-2014 | 4.5 | 1.038 | 0.9936 | 3.1 | 0.879 | 0.9886 | 16.1 | 0.877 | 0.9996 |
| U. Bath | 2010-2013 | 3.9 | 1.086 | 0.9690 | 2.9 | 0.895 | 0.9855 | 15.4 | 0.884 | 0.9987 |
| U. Bath | 2009-2012 | 4.8 | 1.028 | 0.9932 | 3.4 | 0.829 | 0.9972 | 16.3 | 0.870 | 0.9989 |
| Nat. Aut. U. Mexico | 2012-2015 | 4.9 | 1.243 | 0.9924 | 11.4 | 1.101 | 0.9745 | 4.0 | 1.383 | 0.9995 |
| Nat. Aut. U. Mexico | 2011-2014 | 4.8 | 1.243 | 0.9925 | 11.1 | 1.092 | 0.9627 | 3.9 | 1.376 | 0.9986 |
| Nat. Aut. U. Mexico | 2010-2013 | 3.9 | 1.280 | 0.9951 | 9.1 | 1.136 | 0.9822 | 3.9 | 1.374 | 0.9984 |
| Nat. Aut. U. Mexico | 2009-2012 | 4.4 | 1.222 | 0.9722 | 9.0 | 1.123 | 0.9800 | 4.8 | 1.322 | 0.9964 |

Using the Leiden ranking indicators, in the equation
$$N_x = A\, x^\alpha$$
where $N$ is the number of publications in percentile $x$, the two parameters of the power law function can be calculated as
$$A = P_{top1\%}$$
$$\alpha = \lg P_{top10\%} - \lg P_{top1\%}$$
These formulas demonstrate that the number of publications in the 1% most cited papers (e.g., King, 2004) is an imperfect indicator because it does not reveal the real capacity to make an important breakthrough, which is described by smaller percentiles.

**4. Discussion**

This study shows that the frequency distribution of publications on a percentile basis (e.g., Bornmann and Mutz, 2011) is a double rank plot (RNB), which has the property of being well described by a power law. This can be checked by a mathematical approach deriving it from two series of lognormal distributed probability functions, a large series simulating the world publications and a small series simulating country publications. Following this approach, we found that only the upper part of the plot slightly deviated from the power law when the $\mu$ parameter of the simulated country lognormal distribution was very different from the corresponding value in the world distribution.

In comparison to a normal double rank plot, in which the rank number in the country or institution is a function of the world rank number, the percentile-based double rank plot shows lower variability (e.g., compare Fig. 4 to Fig. 3 in RNB). This behavior can be explained by the compression of the scale in the $x$ axes—for the same data, the scale in Fig. 4 varies from 0.5 to 100, while in Fig. 3 in RBN, it varies from 1 to 16,043. Another interesting difference is that the deviation in the upper part of the plot that occasionally occurred in the normal rank plot is much lower in the percentile-based one. Again this might be the effect of the compression although we cannot rule out that the inclusion of the publications with zero citations, which we make in this study but not in the previous one (RNB), also has an effect.

The comparison of research assessments based on the same publications by the two double rank methods—the percentile-based and normal one—were very similar (section 3.2). Certainly, the likelihood of publishing the most cited paper of the year showed certain method dependence. In some cases the results were identical but in others we found variations. Because the results of the two methods should coincide from a mathematical point of view, we assume that differences occur for a more difficult fitting in the normal double rank plot, in which the 30 most cited publications had to be eliminated.

Research assessment based on the upper, power law tail of citation distribution (highly cited papers) is reliable and can be validated in terms of Nobel Prize achievements (Rodríguez-Navarro, 2016). However, it does not have practical application because the power-law tail includes a low proportion of publications (Ruiz-Castillo, 2012; Brzezinski, 2015), which implies that in many research institutions this tail cannot be analyzed. Therefore, we investigated whether the power law tail and the double rank methods produce similar results; Table 2 summarizes the comparison. Attending to the 20-0.2% interval, the results obtained with the percentile-based double rank are absolutely consistent with those based on the analysis of the upper tail. However, on average, the results obtained with the percentile double-rank method were 40% higher than those obtained analyzing the upper tail. In principle, the two methods should give comparable results; and a possible explanation for the 40% deviation using the Nobel Prize percentile and the similarity at the 0.01% percentile comes from the extrapolation required to get such data. While this explanation is found, considering that the results obtained from the upper tail coincide with the garnering of Nobel Prizes (Rodríguez-Navarro, 2016; Rodriguez-Navarro and Narin, 2017), evaluating at the 0.01% percentile level seems to be a reasonably solution. In any case, the percentile-based double rank method allows selecting the percentile by the evaluation agency or institution.

Our results shown in Table 1 were obtained at a high level of aggregation, and through considering only domestic publications. Therefore, our next goal was to test the method at the level of research institutions, applying fractional counting. For this purpose, the data of the Leiden ranking provided an outstanding opportunity. The Leiden indicators P, $P_{top50\%}$, $P_{top10\%}$, and $P_{top1\%}$ are four data points that can be used for the percentile-based double rank analysis. As shown in Sect.3.4, a power law is defined by two parameters, which implies that two data points are sufficient to define a power law. Despite of this, four data points is a low number of points for fitting tests. However, considering that the data points cover two orders of magnitude and that the tests can be repeated 19,000 times with the Leiden ranking data—900 universities, three research fields in natural sciences, and seven time periods—the testing opportunity is really outstanding.

We performed 132 tests with the Leiden data observing good fittings to power laws (Table 3). The $R^2$ values were slightly lower in the most than in the least research-active universities: 0.97 versus 0.99, approximately, which are high values for only four points. Although the low number of data points (Fig. 9) in each case does not allow drawing firm conclusions from the plots, it seems possible that in the most research-active universities, the upper data points (top 100% and perhaps 50%) deviate from the straight line fitted to the other data points, as described in section 3.3 for the USA plots. Apart from these considerations, the analyses of the Leiden indicators demonstrate that real data of research evaluation in institutions using fractional counting behaved exactly as observed in simulations and country analyses.

Although the percentile-based double rank analysis is based on the same properties of the citation distributions as the normal double-rank analysis (RNB), it is much more convenient, independently of a higher statistical robustness that in a general comparison has minor importance. In the first place, the percentile approach is based on a percentile apportionment, which has been extensively investigated (Bornmann, 2010; Bornmann and Mutz, 2011; Leydesdorff and Bornmann, 2011; Leydesdorff et al., 2011; Rousseau, 2012; Bornmann, 2013; Bornmann et al., 2013a; Bornmann et al., 2013b; Waltman and Schreiber, 2013). In fact, the percentile based double rank approach here reported is not a new method of assessment but the application of a mathematical property to an evaluation approach that is extensively used. Thus, while the normal double rank approach requires a specific treatment of the data and fractional counting cannot be performed, the percentile based double rank analysis does not require new data treatments. In addition to many reports, the percentile assessment is used in the Leiden Ranking as already described, SCIMAGO (http://www.scimagoir.com, accessed 01/22/2018), and "Mapping Scientific Excellence" (www.excellencemapping.net/, accessed 01/22/2018). These rankings could include $P_{top\ x\%}$ indicators at low values of $x$ if their authors consider it interesting and this can be done without changing their methods of assessment. Furthermore, from previously published rankings containing the $P_{top10\%}$ and $P_{top1\%}$ indicators, more stringent percentile indicators can be easily obtained. For example, in Table 4 we have added the $P_{top0.1\%}$, $P_{top0.01\%}$, and $P_{top0.001\%}$ indicators to the indicators published previously (Bornmann et al., 2015). The convenience of the use of these more stringent indicators can be ascertained by comparing Switzerland and Spain. Attending to the $P_{top10\%}$, Switzerland and Spain show similar research performances, 49,275 versus 50,797, that no expert would corroborate. In contrast, the $P_{top0.01\%}$ and $P_{top0.001\%}$ indicators suggest a much better performance of Switzerland, which is a much more reasonable assessment. In Table 4, other comparisons are also interesting.

The second and perhaps more important advantage of the percentile based over the normal double rank method is that the arbitrariness of selecting the assessment level can be eliminated. Similarly to the problem of evaluating by highly cited papers (Schreiber, 2013a) the threshold for the normal double rank approach does not have a formal method to fix it. In contrast, in the percentile based approach the percentile can be fixed according to the criterion of experts. For example, to fix the percentile in the topic "electronics" in which there are approximately 6,000 publications in a year, we would ask the experts in electronics how many papers they consider report real breakthroughs in a year. If the answer were six, the right percentile would be 0.1%. The selection of the percentile in this way might have some difficulties because different fields or topics might have different thresholds, but arbitrariness is eliminated.

Table 4. Addition of the $P_{top0.1\%}$, $P_{top0.01\%}$, and $P_{top0.001\%}$ indicators to previously published $P_{top10\%}$, and $P_{top1\%}$ indicators for 30 countries worldwide with the highest percentage of most frequently cited papers (sorted in descending order by $P_{top0.001\%}$)[a]

| Country | $P_{top10\%}$ | $P_{top1\%}$ | $P_{top0.1\%}$ | $P_{top0.01\%}$ | $P_{top0.001\%}$ |
|---|---:|---:|---:|---:|---:|
| US | 858,703 | 96,146 | 10,765 | 1205.33 | 134.96 |
| UK | 201,588 | 20,855 | 2,158 | 223.20 | 23.09 |
| Germany | 159,250 | 15,738 | 1,555 | 153.71 | 15.19 |
| Canada | 100,307 | 10,474 | 1,094 | 114.20 | 11.92 |
| France | 112,965 | 10,971 | 1,065 | 103.48 | 10.05 |
| Switzerland | 49,275 | 5,859 | 697 | 82.84 | 9.85 |
| The Netherlands | 64,667 | 7,060 | 771 | 84.15 | 9.19 |
| Italy | 74,378 | 7,150 | 687 | 66.07 | 6.35 |
| Japan | 109,249 | 9,371 | 804 | 68.95 | 5.91 |
| Australia | 58,612 | 5,854 | 585 | 58.40 | 5.83 |
| China | 75,537 | 6,827 | 617 | 55.77 | 5.04 |
| Sweden | 41,792 | 4,327 | 448 | 46.38 | 4.80 |
| Denmark | 25,022 | 2,832 | 321 | 36.28 | 4.11 |
| Belgium | 29,419 | 3,102 | 327 | 34.49 | 3.64 |
| Spain | 50,797 | 4,526 | 403 | 35.93 | 3.20 |
| Austria | 17,785 | 1,919 | 207 | 22.34 | 2.41 |
| Israel | 22,266 | 2,180 | 213 | 20.90 | 2.05 |
| Finland | 18,247 | 1,837 | 185 | 18.62 | 1.87 |
| Norway | 14,312 | 1,493 | 156 | 16.25 | 1.69 |
| Poland | 12,042 | 1,170 | 114 | 11.04 | 1.07 |
| Korea | 25,233 | 2,037 | 164 | 13.28 | 1.07 |
| New Zealand | 10,361 | 1,026 | 102 | 10.06 | 1.00 |
| Russia | 15,887 | 1,413 | 126 | 11.18 | 0.99 |
| Brazil | 16,025 | 1,309 | 107 | 8.73 | 0.71 |
| Greece | 10,134 | 913 | 82 | 7.41 | 0.67 |
| South Africa | 7,159 | 661 | 61 | 5.64 | 0.52 |
| India | 22,320 | 1,530 | 105 | 7.19 | 0.49 |
| Taiwan | 18,612 | 1,332 | 95 | 6.82 | 0.49 |
| Turkey | 10,100 | 793 | 62 | 4.89 | 0.38 |
| Mexico | 6,169 | 531 | 46 | 3.93 | 0.34 |

[a] The countries and $P_{top10\%}$ and $P_{top1\%}$ indicators have been reproduced from Table 1 in Bormann *et al.* (2015).

The finding that the percentile distribution of publications fits well to a power law across universities or countries reveals the impossibility of describing the research performance by the number of papers in a single percentile, because the results of comparative performances will depend on the selected percentile. This conclusion is graphically shown for simulated institutions in Fig. 2 and for the comparison of the USA and EU in Fig. 6-8. The power law function has two parameters, the coefficient and the exponent. The $P_{top1\%}$ is the coefficient and the lg of the $P_{top10\%}/P_{top1\%}$ ratio is the exponent. This exponent varied from 0.6 to 1.4 (Table 3), which implies very different likelihoods of publishing a very highly cited paper for universities that have the same $P_{top1\%}$ indicator.

It is a fact that in a certain number of universities in middle positions of the Leiden ranking the exponents of the double rank plots are close to 1.0, which implies that the $\mu$ parameters of their lognormal distributions are similar to that of the world distribution. This value implies that the double rank plots are close to straight lines and that the comparative performance is the same or very similar counting the total number of publications or estimating the likelihood at any percentile. In these universities the use of a single percentile for ranking them is correct but this approach cannot be extended to all universities and research institutions.

Unfortunately, we could not find a simple expression for the function that describes the percentile-based double rank plot in all cases. However, this should not be an inconvenience for evaluation, because we found that the power law was an excellent model for percentiles lower than 20%. Although we did not determine the lowest percentile that can be used for university evaluations and ranking attending to statistical considerations, it seems clear that the useful percentile range is sufficiently large. In the three fields of natural sciences of the Leiden ranking, in a significantly high proportion of universities (not less than 50%) the $_{1\%}$ indicator amounts to 5 or less, which might be too low of a value to fit the power law. However, in practically all universities a hypothetic $P_{top5\%}$ indicator would be 8 or higher. Furthermore, the 20% upper limit applies only to top-level universities in which the $P_{top1\%}$ indicator is high. Even if the 20% limit were uniformly applied to all universities and research institutions, the 20-5% percentile interval would allow a reliable calculation of the parameters of the double rank power law.

Finally, current results strengthen our previous results (RNB) showing that the likelihood of making an important discovery can be estimated by the citation distribution of publications that receive a low number of citations. This interdependence of the numbers of highly and lowly cited papers indicates that a research system is a complex system. Therefore, the research policy of a country should aim to improve the proportion of highly versus lowly cited papers, considering, however, that lowly cited papers are in the basis of highly cited papers and cannot be eliminated. Further studies about the bases of the complexity of the research system could help to establish effective country's research policies.

## Acknowledgements

This work was supported by the Spanish Ministerio de Economía y Competitividad, grant numbers FIS2014-52486-R and FIS2017-83709-R.